\documentclass[prl,twocolumn,showpacs,superscriptaddress]{revtex4-1}
\usepackage{graphicx,amssymb,amsmath,psfrag,color}
\usepackage[colorlinks=true,citecolor=blue,linkcolor=blue,urlcolor=blue]{hyperref}
\begin{document}
\definecolor{darkgreen}{rgb}{0,0.5,0}
\newcommand{\be}{\begin{equation}}
\newcommand{\ee}{\end{equation}}
\newcommand{\jav}[1]{\textcolor{red}{#1}}

\title{Momentum-space entanglement and Loschmidt echo in Luttinger liquids after a quantum quench}

\author{Bal\'azs D\'ora}
\email{dora@eik.bme.hu}
\affiliation{Department of Theoretical Physics and BME-MTA Exotic  Quantum  Phases Research Group, Budapest University of Technology and
  Economics, 1521 Budapest, Hungary}
\author{Rex Lundgren}
\email{rexlund@physics.utexas.edu}
\author{Mark Selover}
\affiliation{Department of Physics, The University of Texas at Austin, Austin, Texas 78712, USA}
\author{Frank Pollmann}
\affiliation{Max-Planck-Institut f\"ur Physik komplexer Systeme, 01187 Dresden, Germany}

\date{\today}

\begin{abstract}
Luttinger liquids (LLs) arise by 
coupling left- and right-moving particles through interactions in one dimension. 
This most natural partitioning of LLs is investigated by the momentum-space entanglement 
after a quantum quench using analytical and numerical methods. We show that 
the momentum-space entanglement spectrum of a LL possesses many 
universal features both in equilibrium and after a quantum quench. The largest entanglement 
eigenvalue is identical to the Loschmidt echo, i.e. the overlap of the disentangled and final 
wavefunctions of the system. The second largest eigenvalue is the overlap of the first excited 
state of the disentangled system with zero total momentum and the final wavefunction. The 
entanglement gap is universal both in equilibrium and after a quantum quench. The 
momentum-space entanglement entropy is always extensive and saturates fast to a 
time independent value after the quench, in sharp contrast to a spatial bipartitioning.


\end{abstract}


\maketitle

\paragraph{Introduction. }
There is nothing more quantum mechanical than entanglement, when
the state of certain particles cannot be described independently from the rest.
This constitutes one of the most fundamental differences between classical and quantum physics.
While
entanglement has always been considered as an interesting quantity from its early existence, 
its governing role in various fields of physics
, especially those fields dominated by strong correlations, has only become clear recently \cite{eisert}.
For example, entanglement plays an important role in understanding the thermodynamics of black holes \cite{srednicki}, quantum information theory \cite{nielsen} and (topological) order \cite{amico}.

The entanglement characteristics of a system is obtained usually by partitioning it into two distinct regions, and investigating the properties of the reduced density matrix 
(eigenvalues, entropies etc.) of one of the regions. While this
partitioning is mostly spatial, i.e. done in real-space, other ways of partitioning are equally fruitful. 
In particular, partitioning in \emph{momentum-space} is natural as various instabilities and phase transitions occur
by coupling distinct regions in momentum-space together via interactions.
For example, Cooper pairs are made of particles with
opposite momentum and give rise to superconductivity. 
Density waves are created from  electron-hole pairs with a finite  wavevector  difference.
In one dimension,  Luttinger liquids (LLs), which host many interesting phenomena such as spin-charge separation, charge fractionalization, and non-Fermi liquid behaviour, appear after coupling right- and left-moving fermions together  \cite{giamarchi,nersesyan}. 
Therefore, a momentum-space partition offers a unique perspective on the structure of many-particle wavefunctions~\cite{thomale,qi,lundgrenKK,fuji,lundgren,lundgrenladder,lundgrenspin1,ehlers,Mondragon,Andrade,Balasubramanian,PandoZayas2015,Diptarka,chung}.

Parallel to these developments, non-equilibrium dynamics \cite{polkovnikovrmp,dziarmagareview} have enjoyed immense interest due to experimental advances in cold atomic gases \cite{BlochDalibardZwerger_RMP08}.
In this context, the study of quantum quenches, i.e. time evolving a ground state wavefunction with a different Hamiltonian, is particularly challenging due to the 
interplay of interactions and non-equilibrium dynamics.
Recently proposals have been made about measuring 
entanglement in cold atom systems, e.g., by using 
quantum switches coupled to a composite system consisting of several copies of the original many-body system \cite{Abanin2012} or using quantum interference between the copies \cite{Islam2015}.

In this work, we combine momentum-space entanglement and quantum quenches in a LL. 
We demonstrate that after tracing out the left-movers, 
the entanglement gap (EG), which is the difference between 
the two lowest levels of the entanglement spectrum (ES) \cite{lihaldane}, is universal in a LL \emph{both} in equilibrium
and after a quantum quench and depends only on the Luttinger parameter. 
We show, by using bosonization and numerical exact diagonalization on an interacting lattice model, that 
the largest 
eigenvalue of the reduced density matrix (the lowest level of the ES) is identical to the Loschmidt echo, i.e. 
the overlap of the ground state of the  disentangled, non-interacting system and the final wavefunction
of the interacting system. The Loschmidt echo is related to statistics of work done \cite{rmptalkner,silva} and the Crooks relation \cite{crooks}, 
therefore our results create a direct link between entanglement characteristics and recent developments in non-equilibrium statistical mechanics.

\paragraph{Tomonaga-Luttinger model.} The low energy dynamics of LLs is  described in terms of bosonic sound-like collective excitations, regardless to the  statistics of the original system.
The Hamiltonian of the system is \cite{giamarchi}
\begin{equation}
H=\sum_{q\neq 0}  \omega_q b_q^{\dagger} b^{\phantom{\dagger}}_q
+\frac{g(q)}{2}[b^{\phantom{\dagger}}_qb^{\phantom{\dagger}}_{-q}+b_q^{\dagger} b_{-q}^{\dagger} ],
\label{ham0}
\end{equation}
where $g(q)=g_2|q|$, with $g_2$  the interaction strength, and $\omega_q=v|q|$ the energy of bosonic excitations with momentum $q$.
Starting from the non-interacting limit with $g_2=0$, 
we focus on two extreme cases of turning on interactions, adiabatically and suddenly. In the adiabatic case (which we also refer to as the equilibrium case), the system always remains in its ground state. In the case of a sudden quantum quench, the interaction strength is suddenly changed to some finite value. The final velocity of the system is $v_f=\sqrt{v^2-g_2^2}$.
One can characterize interactions by the dimensionless Luttinger parameter, $K$~\footnote{$K=1$ for the non-interacting case, and $K\gtrless 1$ for attraction/repulsion, respectively.},
 which is given by $K=\sqrt{(v-g_2)/(v+g_2)}$ for Eq. \eqref{ham0}.

The Heisenberg equation of motions, which describe the dynamics of Eq. \eqref{ham0}, couple only pairs of $q$ and $-q$ modes together as
$b_q(t)= u_q(t)b_q +v_q^*(t)b^+_{-q}$,
where all the information about interactions and time dependence is encoded in the Bogoliubov coefficients, $u_q(t)$ and $v_q(t)$, which obey the condition, 
$|u_q^2(t)|-|v_q^2(t)|=1$. Their form is known for the adiabatic and sudden quench case \cite{iucci}.
Since Eq. \eqref{ham0} is quadratic
in bosonic operators, its wavefunction is given by \cite{doraLE}
\begin{gather}
|\Psi(t)\rangle=e^{-i\Phi(t)}
\prod_{q>0}\frac{1}{u_q^*(t)}\exp\left(\frac{v_q^*(t)}{u_q^*(t)} b^{\dagger}_qb^\dagger_{-q}\right)|\Psi_0\rangle,
\end{gather}
with $\Phi(t)$ an overall phase factor and $|\Psi_0\rangle$ is the bosonic vacuum, i.e. the ground state 
wavefunction of the disentangled system.

\paragraph{Entanglement spectrum.} Knowing the explicit form of the wavefunction provides us with the density matrix,
$\rho=|\Psi(t)\rangle\langle \Psi(t)|$.
Instead of calculating the entanglement properties by partitioning our system in real-space, we use the natural partitioning of the wavefunction in terms
of right ($q>0$) and left ($q<0$) moving excitations and trace out all the left-movers. By expanding the exponential in Taylor series in the wavefunction, the reduced
density matrix, $\rho_A(t)$, for the right-movers reads
as
\begin{gather}
\rho_{A}(t)=\exp\left(\sum_{q>0}-\ln|u_q(t)|^2+\ln\frac{|v_q(t)|^2}{|u_q(t)|^2}b^\dagger_qb^{\phantom{\dagger}}_{q}\right).
\end{gather}
The eigenvalues of $\rho_A$
are immediately obtained as
$P_{\{n_q\}}=\prod_{q>0}p_{n_q}(q)$,
 after defining the single particle eigenvalues for a given mode as
$p_n(q)=\left|v_q(t)\right|^{2n}\left|u_q(t)\right|^{-2n-2}$
with $n$ non-negative integer.
The many body ES is obtained as $-\ln(P_{\{n_q\}})$.

The largest eigenvalue of $\rho_A$, whose logarithm is related to the single copy (or $n=\infty$ R\'enyi) entropy \cite{Peschel-2005,Eisert-2005}, is
\begin{gather}
P_{max}=\prod_{q>0}p_0(q)=\prod_{q>0}|u_q(t)|^{-2}.
\label{pmax}
\end{gather}
The
right hand side of Eq. \eqref{pmax} is identified as the Loschmidt echo from Ref. \cite{doraLE} or the return probability, i.e. the overlap of the initial ground state wavefunction
and the final state wavefunction, obtained through the given time evolution, 
as
\begin{gather}
P_{max}=\mathcal L(t)\equiv\left|\langle \Psi_0|\Psi(t)\rangle\right|^2.
\label{pmaxle}
\end{gather}
It creates a direct link between  entanglement, quantum 
quenches \cite{polkovnikovrmp,dziarmagareview} and work statistics \cite{rmptalkner,silva} as the latter is the Fourier transform of $\mathcal L(t)$. This connection and its numerical verification is one of our main results.

The ground state entanglement level of the ES is $-\ln(P_{max})$. The first excited state of the ES, which is defined as the second largest eigenvalue of $\rho_A$, is given by
$P_1=\max_{k>0} \left[p_1(k)\prod_{q>0,q\neq k}p_0(q)\right]$.
The EG (the difference between the ground state and first excited state of the ES) is
$\Delta_{EG}=\ln(P_{max})-\ln(P_1)=2\min_{q>0}\ln\left|{u_q(t)}/{v_q(t)}\right|$,
above which a continuum of many body entanglement levels occurs.

The largest eigenvalue of $\rho_A$ is not the only entanglement level linked to the overlap of two wavefunctions. In fact, all entanglement levels can be linked to 
the overlap of two wavefunctions.
In particular, consider a zero net momentum excited state of the initial wavefunction as  $b^\dagger_{k'}b^\dagger_{-k'}|\Psi_0\rangle$.
Its overlap to the final wavefunction is $p_1(k')\prod_{q>0,q\neq k'}p_0(q)$. Similarly, all other elements of the ES are obtained by
creating net zero momentum boson pairs in the initial wavefunction.

\paragraph{Adiabatic and sudden quench.}
In equilibrium, all Bogoliubov coefficients are time and momentum independent and given by $|v_q(t)|^2=(1-K)^2/4K$.
Therefore, all single particle entanglement eigenvalues, $p_n(q)$, are equal and constant and the ES is flat \cite{fuji,lundgren}. 
The largest eigenvalue is
\begin{gather}
P_{max}=\left[\frac 12+\frac 14\left(K+\frac 1K\right)\right]^{-L/2\pi\alpha}
\label{pmaxad}
\end{gather}
where $1/\alpha$ is the ultraviolet cutoff \cite{giamarchi} and $L$ is the system size.
The EG is universal, i.e. independent of the cutoff, and given by
\begin{gather}
\Delta_{EG}=\ln\left[\frac{(1+K)^2}{(1-K)^2}\right].
\label{egequilibrium}
\end{gather}




After a quantum quench, the Bogoliubov coefficients are time and momentum dependent
and given by $|v_q(t)|^2=\sin^2(v_f|q|t)(1-K^2)^2/4K^2$.
The largest entanglement eigenvalue saturates to the time independent
\begin{gather}
P_{max}=\left[\frac 12+\frac 14\left(K+\frac 1K\right)\right]^{-L/\pi\alpha}
\label{pmaxquench}
\end{gather}
value after a short transient time $\sim \alpha/v$. We refrain from analyzing the full time dependence of $P_{max}$, which has been done in Ref. \cite{doraLE} for the Loschmidt echo.
A finite EG also persists in this case and is given by
\begin{gather}
\Delta_{EG}=\ln\left[\frac{(1+K^2)^2}{(1-K^2)^2}\right].
\label{egquench}
\end{gather}
This EG occurs for modes with $\sin^2(v_f|q|t)=1$. In the thermodynamic limit (TDL), for every time $t$, there is always a momentum state which satisfies this condition.

\begin{figure}[h!]
\centering
\includegraphics[width=7cm]{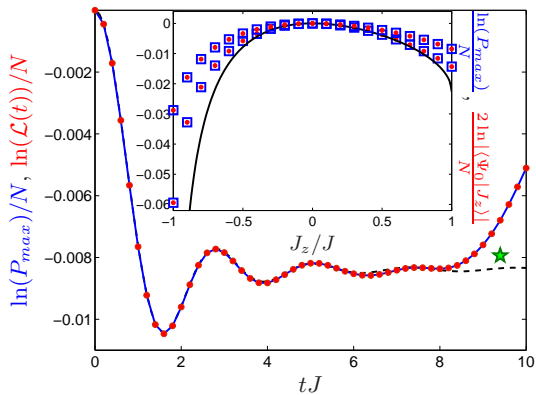}
\caption{(Color online) 
The time evolution of $P_{max}$, Eq.~\eqref{pmaxle}, (blue line) after a quantum quench for the representative case of $J_z=J/2$
and $N=26$ is plotted, together with the overlap of the $J_z=0$ and the quenched
$J_z=J/2$ wavefunction (red circles), falling on top of each other (up to numerical accuracy). The upturn around $tJ=9$ is due to finite size effects, following from the
 comparison of the infinite system result \cite{doraLE} (black dashed line). The green star denotes the prediction of Eq. \eqref{pmaxquench},
which is valid for times much larger than the transient time. The inset visualizes adiabatic results for $P_{max}$ (blue squares) and the overlap of the
 ground state wavefunctions (red dots)
for $N=14$ (upper dataset) and 22 (lower dataset).
The ED data approaches Eq. \eqref{pmaxad} (black line), which is valid in the TDL, upon increasing the system size.}
\label{pmaxfig}
\end{figure}

\paragraph{Exact diagonalization.} 
We now test these analytical predictions numerically using exact diagonalization (ED) on a half-filled tight binding chain of one-dimensional 
spinless fermions with nearest-neighbour interactions and periodic boundary conditions \cite{giamarchi,nersesyan}. 
In momentum-space, the Hamiltonian we consider is 
\begin{gather}
H=J\sum_{k}\cos(k)c_k^{\dagger} c_k^{\phantom{\dagger}}+\frac{J_z}{N}\sum_{k,p,q}\cos(q)c^\dagger_{p-q}c^{\phantom{\dagger}}_pc^\dagger_{k+q}c^{\phantom{\dagger}}_k,
\label{hmom}
\end{gather}
where $c$'s are fermionic annihilation operators in momentum-space, $N$ the number of lattice sites and $k=2\pi m/N$, $n=1\dots N$. Its low energy physics is accounted for by Eq. \eqref{ham0} with
 $K=\pi/2[\pi-\arccos(J_z/J)]$, covering $1/2 (J_z=J)<K<\infty (J_z=-J)$ \cite{giamarchi}. \footnote{ By a  
momentum-space Jordan-Wigner transformation, we arrive at a momentum-space Hamiltonian of hard-core bosons,  which is solved numerically using ED or the Lanczos method.} 
We consider system sizes of $N=10$, 14, 18, 22 and 26 (to avoid a degenerate Fermi sea) and then perform finite size scaling to reach the TDL. For the range of parameters we consider, there is a unique non-degenerate ground state. 
We compute the time evolution of the momentum-space wavefunction using the Krylov method until $tJ=20$ \cite{Saad-1992}.
After constructing the density matrix from the wavefunction in momentum-space basis, all left-movers (i.e. $k<0$ modes) are traced out, giving the reduced density matrix.

\begin{figure}[h!]
\centering
\includegraphics[width=7cm]{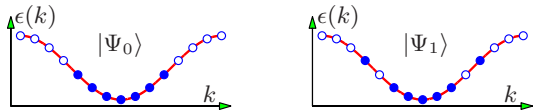}
\caption{(Color online) Schematic picture of the ground (left) and first excited (right) state of the half filled tight binding model with zero total momentum
  is visualized in momentum-space ($\epsilon(k)=J\cos(k)$),
whose overlap with the final wavefunction yields the first two
entanglement eigenvalues.}
\label{overlapspec}
\end{figure}

Eqs. \eqref{pmax} and \eqref{pmaxle} predict that the largest eigenvalue of $\rho_A$ is equal to the overlap of the decoupled and coupled wavefunction.
Using ED, $P_{max}$, as well as the overlap of the $J_z=0$ ground state wavefunction with the finite $J_z$ adiabatic or sudden quench
wavefunction, are evaluated and shown in Fig. \ref{pmaxfig}.
These agree perfectly both in equilibrium and after the quench (within numerical accuracy), thus confirming the bosonization prediction.
The  $L/\pi\alpha$ factor in  Eqs.  \eqref{pmaxad} and \eqref{pmaxquench}
contains the short distance cutoff. 
The short distance cutoff is estimated from the fidelity
susceptibility $\chi_f$ \cite{doraLE}, which gives $L/2\pi\alpha\approx N\chi_f\pi^2$ with  $\chi_f\approx 0.0195$ \cite{sirker}.
Eq. \eqref{pmaxle} also works at the  $J_z=-J$ point (not shown), which falls outside of the validity of bosonization because the excitation spectrum
changes from linear to quadratic in momentum. 
We conjecture that Eq. \eqref{pmaxle} is valid for other systems as well. 
We have numerically checked this conjecture on other systems \cite{EPAPS}, including the non-integrable extension 
(by adding second nearest neighbour hopping \cite{mukerjee,bartsch}) of Eq.~\eqref{hmom}, and different boundary 
conditions with affirmative conclusion.

From the analysis of the the numerical data (not shown), we find that for all system sizes considered and within the critical region, the second largest eigenvalue of $\rho_A$ equals the overlap between the final, coupled wavefunction and the first excited state of the disentangled wavefunction with total momentum $0$, arising from annihilating
a left and a right mover from right below the filled Fermi sea and creating a left and right mover right above the Fermi sea, shown
in Fig. \ref{overlapspec}, i.e.
$P_1=|\langle \Psi_1|\Psi(t)\rangle|^2$. 
The numerical data for $P_1$ is in excellent agreement with the bosonization prediction.
In the case of a quantum quench, $P_1$ (during its time evolution) is equal to the
overlap between the time evolved wavefunction and a general total momentum zero excited state of the initial disentangled Hamiltonian, but \emph{not necessarily} the first zero momentum excited state. However, the minimal value of $P_1$ over time, which defines the EG, occurs for the first zero momentum excited state.

\begin{figure}[h!]
\centering
\includegraphics[width=6.6cm]{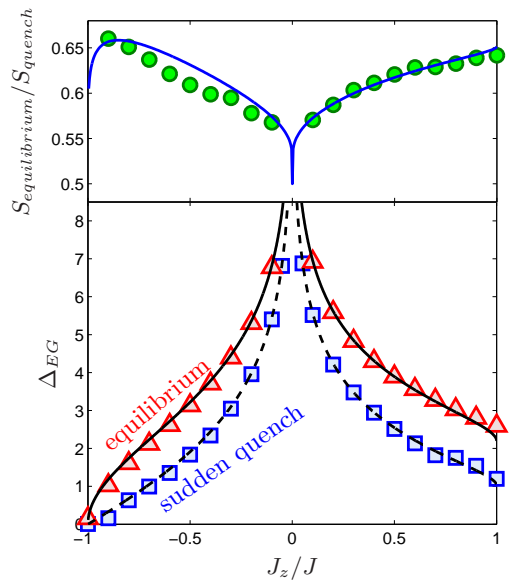}
\caption{(Color online) Lower panel: The EG extrapolated to the TDL in equilibrium (red triangles) and after
a quench (blue squares). 
The black solid and dashed lines depict the analytical
results without any fitting parameter from Eqs. \eqref{egequilibrium}
and \eqref{egquench}, respectively.
Upper panel: The ratio (circles) of the von-Neumann  entanglement entropies in the steady state after the quench and in equilibrium from ED, extrapolated to the TDL, compared to
the Luttinger model result (solid line)  from Eq. \eqref{shannonent}, without any fitting.}
\label{EGscaling}
\end{figure}

The EG, which is the minimal difference between the ground state and first excited state of the ES, is shown in Fig. \ref{EGscaling}. 
After a quantum quench, the EG is obtained from the difference of the first two eigenvalues of $\rho_A$ after the transient time but before the revival time $N/J$, 
when finite size effects become important. The EG is largely insensitive to variations 
within this window.  The same results are obtained when the EG is calculated from the overlaps
of the final wavefunction with the states shown in Fig. \ref{overlapspec}.
Around the non-interacting XX point, $\Delta_{EG}\sim -\ln(J_z^2)$ since
the left and right moving fermions are perfectly  disentangled in this case.
Both in equilibrium and after the quantum quench, the EG is universal and in near perfect agreement with bosonization prediction throughout the critical region, depending only on $K$. The EG stays finite for $J_z>0$ within the LL phase, and does not vanish even at the BKT transition point at $J_z=J$. On the other hand, the EG collapses at $J_z=-J$, signaling the first order quantum critical point. 
In equilibrium, $\Delta_{EG}\sim \sqrt{J_z+J}$ for $J_z\gtrsim -J$, which changes to $\Delta_{EG}\sim J_z+J$ after a quantum quench. We suspect that the EG is sensitive to any finite order quantum critical point, except possibly for the BKT transition, being of infinite order. The detection of the BKT transition is also difficult using the fidelity susceptibility \cite{sirker}. One possible scenario is that the EG does not detect the BKT transition due to the non-universality of the ES \cite{chandran}. In fact, recent numerical work on the momentum-space ES of the XXZ spin-half chain strongly suggests that EG (in equilibrium) remains open in the gapped phase \cite{lundgren}. We leave the non-universality of the ES after a quantum quench an open question.

\paragraph{Entanglement entropy.} The amount of entanglement is quantified by the (von-Neumann or R\'enyi) entanglement entropy.  
The momentum-space von-Neumann entanglement entropy is
\begin{gather}
S=2\sum_{q>0}|u_q(t)|^2\ln\left|u_q(t)\right|-|v_q(t)|^2\ln\left|v_q(t)\right|
\label{shannonent}
\end{gather}
from bosonization, and the Bogoliubov coefficients encode the properties of the final wavefunction. 
For a spatial bipartition, 
$S\sim \ln(N)$ and $\sim t$ in equilibrium and after a quench for a LL, respectively  \cite{eisert}.
For the present case, interactions connects $q$ and $-q$ states, and by halving
momentum-space for partitioning, the number of bonds we cut through is proportional to the number of momentum 
states, i.e. $\sim N$. Thus, a volume law holds for the entropy both in equilibrium and after the quench. 
By inspecting Eq. \eqref{shannonent}, $S$ grows as $\sim Lt^2\ln(1/t)$ initially after the quench, before 
 saturating (within a timescale $\alpha/v$ or $1/J$) to a finite, $\sim L$ but  $t$ independent value in the TDL.
These results are reproduced by ED.
Fig. \eqref{EGscaling} shows the ratio of the entropies (in steady state) after a quantum quench and in equilibrium from ED, free from the extensive 
$N$ or $L/\alpha$ prefactor. We see that $S_{quench}>S_{equilibrium}$ (after a short initial transient time) and
their biggest ratio is 2. This ratio is reached as one approaches the non-interacting limit. 

\paragraph{Discussion.}
We have demonstrated that the low lying entanglement levels of LLs are identical to the overlap of the final wavefunction and the zero total momentum 
eigenstates of the disentangled Hamiltonian.
The EG persists throughout the critical phase and is a universal function of the Luttinger parameter both in equilibrium and after a quantum quench.
Our results create a unique link between entanglement and the Loschmidt echo \cite{silva} as well as the closely related work statistics \cite{rmptalkner}. We note our approach also quantifies the 
entanglement between the up and down spins in the spin filtered helical edge state of a quantum spin-Hall insulator \cite{cwu}.

The momentum-space entanglement entropy is always extensive, and grows initially as $S\sim Lt^2\ln(1/t)$ after a quantum quench, before saturating to a time independent value.
Understanding momentum-space entanglement bears relevance for the development of momentum-space numerical techniques built on optimally entangled states \cite{ehlers,motruk}.
The similar magnitude of the entanglement entropies in equilibrium and after quench
as well as the separation of entanglement scales, i.e. the EG, near the critical point could help improve the density matrix renormalization group algorithm \cite{ehlers,lundgren}.
For example, the EG seems to stay finite at the BKT transition point even after the quench, facilitating numerics \cite{lundgren}.
Our work also present alternatives to measuring the entanglement\cite{Islam2015} through the Loschmidt echo.

\begin{acknowledgments}

We are grateful to Mikl\'os A. Werner for discussions on ED and for the computing resources in the MPI-PKS and CPFS in Dresden. B.D. is supported by the Hungarian Scientific  Research Funds Nos. K101244, K105149, K108676
and by the Bolyai Program of the HAS.
R.L. is supported by National Science Foundation Graduate Research Fellowship award number 2012115499.
\end{acknowledgments}

\bibliographystyle{apsrev}
\bibliography{wboson1}

\begin{thebibliography}{10}
\expandafter\ifx\csname bibnamefont\endcsname\relax
  \def\bibnamefont#1{#1}\fi
\expandafter\ifx\csname bibfnamefont\endcsname\relax
  \def\bibfnamefont#1{#1}\fi
\expandafter\ifx\csname url\endcsname\relax
  \def\url#1{\texttt{#1}}\fi
\expandafter\ifx\csname urlprefix\endcsname\relax\def\urlprefix{URL }\fi
\providecommand{\bibinfo}[2]{#2}
\providecommand{\eprint}[2][]{\url{#2}}

\bibitem{eisert}
\bibinfo{author}{\bibfnamefont{J.}~\bibnamefont{Eisert}},
  \bibinfo{author}{\bibfnamefont{M.}~\bibnamefont{Cramer}}, \bibnamefont{and}
  \bibinfo{author}{\bibfnamefont{M.~B.} \bibnamefont{Plenio}},
  \emph{\bibinfo{title}{\textit{Colloquium} : Area laws for the entanglement
  entropy}}, \bibinfo{journal}{Rev. Mod. Phys.} \textbf{\bibinfo{volume}{82}},
  \bibinfo{pages}{277} (\bibinfo{year}{2010}).

\bibitem{srednicki}
\bibinfo{author}{\bibfnamefont{M.}~\bibnamefont{Srednicki}},
  \emph{\bibinfo{title}{Entropy and area}}, \bibinfo{journal}{Phys. Rev. Lett.}
  \textbf{\bibinfo{volume}{71}}, \bibinfo{pages}{666} (\bibinfo{year}{1993}).

\bibitem{nielsen}
\bibinfo{author}{\bibfnamefont{M.}~\bibnamefont{Nielsen}} \bibnamefont{and}
  \bibinfo{author}{\bibfnamefont{I.}~\bibnamefont{Chuang}},
  \emph{\bibinfo{title}{Quantum Computation and Quantum Information}}
  (\bibinfo{publisher}{Cambridge University Press},
  \bibinfo{address}{Cambridge}, \bibinfo{year}{2000}).

\bibitem{amico}
\bibinfo{author}{\bibfnamefont{L.}~\bibnamefont{Amico}},
  \bibinfo{author}{\bibfnamefont{R.}~\bibnamefont{Fazio}},
  \bibinfo{author}{\bibfnamefont{A.}~\bibnamefont{Osterloh}}, \bibnamefont{and}
  \bibinfo{author}{\bibfnamefont{V.}~\bibnamefont{Vedral}},
  \emph{\bibinfo{title}{Entanglement in many-body systems}},
  \bibinfo{journal}{Rev. Mod. Phys.} \textbf{\bibinfo{volume}{80}},
  \bibinfo{pages}{517} (\bibinfo{year}{2008}).

\bibitem{giamarchi}
\bibinfo{author}{\bibfnamefont{T.}~\bibnamefont{Giamarchi}},
  \emph{\bibinfo{title}{Quantum Physics in One Dimension}}
  (\bibinfo{publisher}{Oxford University Press}, \bibinfo{address}{Oxford},
  \bibinfo{year}{2004}).

\bibitem{nersesyan}
\bibinfo{author}{\bibfnamefont{A.~O.} \bibnamefont{Gogolin}},
  \bibinfo{author}{\bibfnamefont{A.~A.} \bibnamefont{Nersesyan}},
  \bibnamefont{and} \bibinfo{author}{\bibfnamefont{A.~M.}
  \bibnamefont{Tsvelik}}, \emph{\bibinfo{title}{Bosonization and Strongly
  Correlated Systems}} (\bibinfo{publisher}{Cambridge University Press},
  \bibinfo{address}{Cambridge}, \bibinfo{year}{1998}).

\bibitem{thomale}
\bibinfo{author}{\bibfnamefont{R.}~\bibnamefont{Thomale}},
  \bibinfo{author}{\bibfnamefont{D.~P.} \bibnamefont{Arovas}},
  \bibnamefont{and} \bibinfo{author}{\bibfnamefont{B.~A.}
  \bibnamefont{Bernevig}}, \emph{\bibinfo{title}{Nonlocal order in gapless
  systems: Entanglement spectrum in spin chains}}, \bibinfo{journal}{Phys. Rev.
  Lett.} \textbf{\bibinfo{volume}{105}}, \bibinfo{pages}{116805}
  (\bibinfo{year}{2010}).

\bibitem{qi}
\bibinfo{author}{\bibfnamefont{X.-L.} \bibnamefont{Qi}},
  \bibinfo{author}{\bibfnamefont{H.}~\bibnamefont{Katsura}}, \bibnamefont{and}
  \bibinfo{author}{\bibfnamefont{A.~W.~W.} \bibnamefont{Ludwig}},
  \emph{\bibinfo{title}{General relationship between the entanglement spectrum
  and the edge state spectrum of topological quantum states}},
  \bibinfo{journal}{Phys. Rev. Lett.} \textbf{\bibinfo{volume}{108}},
  \bibinfo{pages}{196402} (\bibinfo{year}{2012}).

\bibitem{lundgrenKK}
\bibinfo{author}{\bibfnamefont{R.}~\bibnamefont{Lundgren}},
  \bibinfo{author}{\bibfnamefont{V.}~\bibnamefont{Chua}}, \bibnamefont{and}
  \bibinfo{author}{\bibfnamefont{G.~A.} \bibnamefont{Fiete}},
  \emph{\bibinfo{title}{Entanglement entropy and spectra of the one-dimensional
  kugel-khomskii model}}, \bibinfo{journal}{Phys. Rev. B}
  \textbf{\bibinfo{volume}{86}}, \bibinfo{pages}{224422}
  (\bibinfo{year}{2012}).

\bibitem{fuji}
\bibinfo{author}{\bibfnamefont{R.}~\bibnamefont{Lundgren}},
  \bibinfo{author}{\bibfnamefont{Y.}~\bibnamefont{Fuji}},
  \bibinfo{author}{\bibfnamefont{S.}~\bibnamefont{Furukawa}}, \bibnamefont{and}
  \bibinfo{author}{\bibfnamefont{M.}~\bibnamefont{Oshikawa}},
  \emph{\bibinfo{title}{Entanglement spectra between coupled tomonaga-luttinger
  liquids: Applications to ladder systems and topological phases}},
  \bibinfo{journal}{Phys. Rev. B} \textbf{\bibinfo{volume}{88}},
  \bibinfo{pages}{245137} (\bibinfo{year}{2013}).

\bibitem{lundgren}
\bibinfo{author}{\bibfnamefont{R.}~\bibnamefont{Lundgren}},
  \bibinfo{author}{\bibfnamefont{J.}~\bibnamefont{Blair}},
  \bibinfo{author}{\bibfnamefont{M.}~\bibnamefont{Greiter}},
  \bibinfo{author}{\bibfnamefont{A.}~\bibnamefont{L\"auchli}},
  \bibinfo{author}{\bibfnamefont{G.~A.} \bibnamefont{Fiete}}, \bibnamefont{and}
  \bibinfo{author}{\bibfnamefont{R.}~\bibnamefont{Thomale}},
  \emph{\bibinfo{title}{Momentum-space entanglement spectrum of bosons and
  fermions with interactions}}, \bibinfo{journal}{Phys. Rev. Lett.}
  \textbf{\bibinfo{volume}{113}}, \bibinfo{pages}{256404}
  (\bibinfo{year}{2014}).

\bibitem{lundgrenladder}
\bibinfo{author}{\bibfnamefont{R.}~\bibnamefont{Lundgren}},
  \emph{\bibinfo{title}{Momentum-space entanglement in heisenberg spin-half
  ladders}}, \bibinfo{journal}{Phys. Rev. B} \textbf{\bibinfo{volume}{93}},
  \bibinfo{pages}{125107} (\bibinfo{year}{2016}).

\bibitem{lundgrenspin1}
\bibinfo{author}{\bibfnamefont{R.}~\bibnamefont{{Lundgren}}},
  \bibinfo{author}{\bibfnamefont{J.}~\bibnamefont{{Blair}}},
  \bibinfo{author}{\bibfnamefont{P.}~\bibnamefont{{Laurell}}},
  \bibinfo{author}{\bibfnamefont{N.}~\bibnamefont{{Regnault}}},
  \bibinfo{author}{\bibfnamefont{G.~A.} \bibnamefont{{Fiete}}},
  \bibinfo{author}{\bibfnamefont{M.}~\bibnamefont{{Greiter}}},
  \bibnamefont{and}
  \bibinfo{author}{\bibfnamefont{R.}~\bibnamefont{{Thomale}}},
  \emph{\bibinfo{title}{{Universal entanglement spectra in critical spin
  chains}}}, \bibinfo{note}{arXiv:1512.09030}.

\bibitem{ehlers}
\bibinfo{author}{\bibfnamefont{G.}~\bibnamefont{Ehlers}},
  \bibinfo{author}{\bibfnamefont{J.}~\bibnamefont{S\'olyom}},
  \bibinfo{author}{\bibfnamefont{O.}~\bibnamefont{Legeza}}, \bibnamefont{and}
  \bibinfo{author}{\bibfnamefont{R.~M.} \bibnamefont{Noack}},
  \emph{\bibinfo{title}{Entanglement structure of the hubbard model in momentum
  space}}, \bibinfo{journal}{Phys. Rev. B} \textbf{\bibinfo{volume}{92}},
  \bibinfo{pages}{235116} (\bibinfo{year}{2015}).

\bibitem{Mondragon}
\bibinfo{author}{\bibfnamefont{I.}~\bibnamefont{Mondragon-Shem}},
  \bibinfo{author}{\bibfnamefont{M.}~\bibnamefont{Khan}}, \bibnamefont{and}
  \bibinfo{author}{\bibfnamefont{T.~L.} \bibnamefont{Hughes}},
  \emph{\bibinfo{title}{Characterizing disordered fermion systems using the
  momentum-space entanglement spectrum}}, \bibinfo{journal}{Phys. Rev. Lett.}
  \textbf{\bibinfo{volume}{110}}, \bibinfo{pages}{046806}
  (\bibinfo{year}{2013}).

\bibitem{Andrade}
\bibinfo{author}{\bibfnamefont{E.~C.} \bibnamefont{Andrade}},
  \bibinfo{author}{\bibfnamefont{M.}~\bibnamefont{Steudtner}},
  \bibnamefont{and} \bibinfo{author}{\bibfnamefont{M.}~\bibnamefont{Vojta}},
  \emph{\bibinfo{title}{Anderson localization and momentum-space
  entanglement}}, \bibinfo{journal}{Journal of Statistical Mechanics: Theory
  and Experiment} \textbf{\bibinfo{volume}{2014}}, \bibinfo{pages}{P07022}
  (\bibinfo{year}{2014}).

\bibitem{Balasubramanian}
\bibinfo{author}{\bibfnamefont{V.}~\bibnamefont{Balasubramanian}},
  \bibinfo{author}{\bibfnamefont{M.~B.} \bibnamefont{McDermott}},
  \bibnamefont{and}
  \bibinfo{author}{\bibfnamefont{M.}~\bibnamefont{Van~Raamsdonk}},
  \emph{\bibinfo{title}{Momentum-space entanglement and renormalization in
  quantum field theory}}, \bibinfo{journal}{Phys. Rev. D}
  \textbf{\bibinfo{volume}{86}}, \bibinfo{pages}{045014}
  (\bibinfo{year}{2012}).

\bibitem{PandoZayas2015}
\bibinfo{author}{\bibfnamefont{L.~A.} \bibnamefont{Pando~Zayas}}
  \bibnamefont{and} \bibinfo{author}{\bibfnamefont{N.}~\bibnamefont{Quiroz}},
  \emph{\bibinfo{title}{Left-right entanglement entropy of boundary states}},
  \bibinfo{journal}{Journal of High Energy Physics}
  \textbf{\bibinfo{volume}{2015}}, \bibinfo{pages}{1} (\bibinfo{year}{2015}).

\bibitem{Diptarka}
\bibinfo{author}{\bibfnamefont{D.}~\bibnamefont{Das}} \bibnamefont{and}
  \bibinfo{author}{\bibfnamefont{S.}~\bibnamefont{Datta}},
  \emph{\bibinfo{title}{Universal features of left-right entanglement
  entropy}}, \bibinfo{journal}{Phys. Rev. Lett.}
  \textbf{\bibinfo{volume}{115}}, \bibinfo{pages}{131602}
  (\bibinfo{year}{2015}).

\bibitem{chung}
\bibinfo{author}{\bibfnamefont{M.-C.} \bibnamefont{Chung}},
  \bibinfo{author}{\bibfnamefont{A.}~\bibnamefont{Iucci}}, \bibnamefont{and}
  \bibinfo{author}{\bibfnamefont{M.~A.} \bibnamefont{Cazalilla}},
  \emph{\bibinfo{title}{Thermalization in systems with bipartite eigenmode
  entanglement}}, \bibinfo{journal}{New J. Phys.}
  \textbf{\bibinfo{volume}{14}}, \bibinfo{pages}{075013}
  (\bibinfo{year}{2012}).

\bibitem{polkovnikovrmp}
\bibinfo{author}{\bibfnamefont{A.}~\bibnamefont{Polkovnikov}},
  \bibinfo{author}{\bibfnamefont{K.}~\bibnamefont{Sengupta}},
  \bibinfo{author}{\bibfnamefont{A.}~\bibnamefont{Silva}}, \bibnamefont{and}
  \bibinfo{author}{\bibfnamefont{M.}~\bibnamefont{Vengalattore}},
  \emph{\bibinfo{title}{\textit{Colloquium} : Nonequilibrium dynamics of closed
  interacting quantum systems}}, \bibinfo{journal}{Rev. Mod. Phys.}
  \textbf{\bibinfo{volume}{83}}, \bibinfo{pages}{863} (\bibinfo{year}{2011}).

\bibitem{dziarmagareview}
\bibinfo{author}{\bibfnamefont{J.}~\bibnamefont{Dziarmaga}},
  \emph{\bibinfo{title}{Dynamics of a quantum phase transition and relaxation
  to a steady state}}, \bibinfo{journal}{Adv. Phys.}
  \textbf{\bibinfo{volume}{59}}, \bibinfo{pages}{1063} (\bibinfo{year}{2010}).

\bibitem{BlochDalibardZwerger_RMP08}
\bibinfo{author}{\bibfnamefont{I.}~\bibnamefont{Bloch}},
  \bibinfo{author}{\bibfnamefont{J.}~\bibnamefont{Dalibard}}, \bibnamefont{and}
  \bibinfo{author}{\bibfnamefont{W.}~\bibnamefont{Zwerger}},
  \emph{\bibinfo{title}{Many-body physics with ultracold gases}},
  \bibinfo{journal}{Rev. Mod. Phys.} \textbf{\bibinfo{volume}{80}},
  \bibinfo{pages}{885} (\bibinfo{year}{2008}).

\bibitem{Abanin2012}
\bibinfo{author}{\bibfnamefont{D.~A.} \bibnamefont{Abanin}} \bibnamefont{and}
  \bibinfo{author}{\bibfnamefont{E.}~\bibnamefont{Demler}},
  \emph{\bibinfo{title}{Measuring entanglement entropy of a generic many-body
  system with a quantum switch}}, \bibinfo{journal}{Phys. Rev. Lett.}
  \textbf{\bibinfo{volume}{109}}, \bibinfo{pages}{020504}
  (\bibinfo{year}{2012}).

\bibitem{Islam2015}
\bibinfo{author}{\bibfnamefont{R.}~\bibnamefont{Islam}},
  \bibinfo{author}{\bibfnamefont{R.}~\bibnamefont{Ma}},
  \bibinfo{author}{\bibfnamefont{P.~M.} \bibnamefont{Preiss}},
  \bibinfo{author}{\bibfnamefont{M.~E.} \bibnamefont{Tai}},
  \bibinfo{author}{\bibfnamefont{A.}~\bibnamefont{Lukin}},
  \bibinfo{author}{\bibfnamefont{M.}~\bibnamefont{Rispoli}}, \bibnamefont{and}
  \bibinfo{author}{\bibfnamefont{M.}~\bibnamefont{Greiner}},
  \emph{\bibinfo{title}{Measuring entanglement entropy in a quantum many-body
  system}}, \bibinfo{journal}{Nature} \textbf{\bibinfo{volume}{528}},
  \bibinfo{pages}{77} (\bibinfo{year}{2015}).

\bibitem{lihaldane}
\bibinfo{author}{\bibfnamefont{H.}~\bibnamefont{Li}} \bibnamefont{and}
  \bibinfo{author}{\bibfnamefont{F.~D.~M.} \bibnamefont{Haldane}},
  \emph{\bibinfo{title}{Entanglement spectrum as a generalization of
  entanglement entropy: Identification of topological order in non-abelian
  fractional quantum hall effect states}}, \bibinfo{journal}{Phys. Rev. Lett.}
  \textbf{\bibinfo{volume}{101}}, \bibinfo{pages}{010504}
  (\bibinfo{year}{2008}).

\bibitem{rmptalkner}
\bibinfo{author}{\bibfnamefont{M.}~\bibnamefont{Campisi}},
  \bibinfo{author}{\bibfnamefont{P.}~\bibnamefont{H\"anggi}}, \bibnamefont{and}
  \bibinfo{author}{\bibfnamefont{P.}~\bibnamefont{Talkner}},
  \emph{\bibinfo{title}{\textit{Colloquium} : Quantum fluctuation relations:
  Foundations and applications}}, \bibinfo{journal}{Rev. Mod. Phys.}
  \textbf{\bibinfo{volume}{83}}, \bibinfo{pages}{771} (\bibinfo{year}{2011}).

\bibitem{silva}
\bibinfo{author}{\bibfnamefont{A.}~\bibnamefont{Silva}},
  \emph{\bibinfo{title}{Statistics of the work done on a quantum critical
  system by quenching a control parameter}}, \bibinfo{journal}{Phys. Rev.
  Lett.} \textbf{\bibinfo{volume}{101}}, \bibinfo{pages}{120603}
  (\bibinfo{year}{2008}).

\bibitem{crooks}
\bibinfo{author}{\bibfnamefont{G.~E.} \bibnamefont{Crooks}},
  \emph{\bibinfo{title}{Entropy production fluctuation theorem and the
  nonequilibrium work relation for free energy differences}},
  \bibinfo{journal}{Phys. Rev. E} \textbf{\bibinfo{volume}{60}},
  \bibinfo{pages}{2721} (\bibinfo{year}{1999}).

\bibitem{Note1}
\bibinfo{note}{$K=1$ for the non-interacting case, and $K\gtrless 1$ for
  attraction/repulsion, respectively.}

\bibitem{iucci}
\bibinfo{author}{\bibfnamefont{A.}~\bibnamefont{Iucci}} \bibnamefont{and}
  \bibinfo{author}{\bibfnamefont{M.~A.} \bibnamefont{Cazalilla}},
  \emph{\bibinfo{title}{Quantum quench dynamics of the luttinger model}},
  \bibinfo{journal}{Phys. Rev. A} \textbf{\bibinfo{volume}{80}},
  \bibinfo{pages}{063619} (\bibinfo{year}{2009}).

\bibitem{doraLE}
\bibinfo{author}{\bibfnamefont{B.}~\bibnamefont{D\'ora}},
  \bibinfo{author}{\bibfnamefont{F.}~\bibnamefont{Pollmann}},
  \bibinfo{author}{\bibfnamefont{J.}~\bibnamefont{Fort\'agh}},
  \bibnamefont{and} \bibinfo{author}{\bibfnamefont{G.}~\bibnamefont{Zar\'and}},
  \emph{\bibinfo{title}{Loschmidt echo and the many-body orthogonality
  catastrophe in a qubit-coupled luttinger liquid}}, \bibinfo{journal}{Phys.
  Rev. Lett.} \textbf{\bibinfo{volume}{111}}, \bibinfo{pages}{046402}
  (\bibinfo{year}{2013}).

\bibitem{Peschel-2005}
\bibinfo{author}{\bibfnamefont{I.}~\bibnamefont{Peschel}} \bibnamefont{and}
  \bibinfo{author}{\bibfnamefont{J.}~\bibnamefont{Zhao}},
  \emph{\bibinfo{title}{On single-copy entanglement}}, \bibinfo{journal}{J.
  Stat. Mech.} p. \bibinfo{pages}{P11002} (\bibinfo{year}{2005}).

\bibitem{Eisert-2005}
\bibinfo{author}{\bibfnamefont{J.}~\bibnamefont{Eisert}} \bibnamefont{and}
  \bibinfo{author}{\bibfnamefont{M.}~\bibnamefont{Cramer}},
  \emph{\bibinfo{title}{Single-copy entanglement in critical quantum spin
  chains}}, \bibinfo{journal}{Phys. Rev. A} \textbf{\bibinfo{volume}{72}},
  \bibinfo{pages}{042112} (\bibinfo{year}{2005}).

\bibitem{Note2}
\bibinfo{note}{By a momentum-space Jordan-Wigner transformation, we arrive at a
  momentum-space Hamiltonian of hard-core bosons, which is solved numerically
  using ED or the Lanczos method.}

\bibitem{Saad-1992}
\bibinfo{author}{\bibfnamefont{Y.}~\bibnamefont{Saad}},
  \emph{\bibinfo{title}{Analysis of some krylov subspace approximations to the
  matrix exponential operator}}, \bibinfo{journal}{SIAM Journal on Numerical
  Analysis} \textbf{\bibinfo{volume}{29}}, \bibinfo{pages}{209}
  (\bibinfo{year}{1992}).

\bibitem{sirker}
\bibinfo{author}{\bibfnamefont{J.}~\bibnamefont{Sirker}},
  \emph{\bibinfo{title}{Finite-temperature fidelity susceptibility for
  one-dimensional quantum systems}}, \bibinfo{journal}{Phys. Rev. Lett.}
  \textbf{\bibinfo{volume}{105}}, \bibinfo{pages}{117203}
  (\bibinfo{year}{2010}).

\bibitem{EPAPS}
\bibinfo{note}{See EPAPS Document No. XXX for supplementary material providing
  further analysis on other models.}

\bibitem{mukerjee}
\bibinfo{author}{\bibfnamefont{S.}~\bibnamefont{Mukerjee}} \bibnamefont{and}
  \bibinfo{author}{\bibfnamefont{B.~S.} \bibnamefont{Shastry}},
  \emph{\bibinfo{title}{Signatures of diffusion and ballistic transport in the
  stiffness, dynamical correlation functions, and statistics of one-dimensional
  systems}}, \bibinfo{journal}{Phys. Rev. B} \textbf{\bibinfo{volume}{77}},
  \bibinfo{pages}{245131} (\bibinfo{year}{2008}).

\bibitem{bartsch}
\bibinfo{author}{\bibfnamefont{C.}~\bibnamefont{Bartsch}} \bibnamefont{and}
  \bibinfo{author}{\bibfnamefont{J.}~\bibnamefont{Gemmer}},
  \emph{\bibinfo{title}{Boltzmann-type approach to transport in weakly
  interacting one-dimensional fermionic systems}}, \bibinfo{journal}{Phys. Rev.
  E} \textbf{\bibinfo{volume}{85}}, \bibinfo{pages}{041103}
  (\bibinfo{year}{2012}).

\bibitem{chandran}
\bibinfo{author}{\bibfnamefont{A.}~\bibnamefont{Chandran}},
  \bibinfo{author}{\bibfnamefont{V.}~\bibnamefont{Khemani}}, \bibnamefont{and}
  \bibinfo{author}{\bibfnamefont{S.~L.} \bibnamefont{Sondhi}},
  \emph{\bibinfo{title}{How universal is the entanglement spectrum?}},
  \bibinfo{journal}{Phys. Rev. Lett.} \textbf{\bibinfo{volume}{113}},
  \bibinfo{pages}{060501} (\bibinfo{year}{2014}).

\bibitem{cwu}
\bibinfo{author}{\bibfnamefont{C.}~\bibnamefont{Wu}},
  \bibinfo{author}{\bibfnamefont{B.~A.} \bibnamefont{Bernevig}},
  \bibnamefont{and} \bibinfo{author}{\bibfnamefont{S.-C.} \bibnamefont{Zhang}},
  \emph{\bibinfo{title}{Helical liquid and the edge of quantum spin hall
  systems}}, \bibinfo{journal}{Phys. Rev. Lett.} \textbf{\bibinfo{volume}{96}},
  \bibinfo{pages}{106401} (\bibinfo{year}{2006}).

\bibitem{motruk}
\bibinfo{author}{\bibfnamefont{J.}~\bibnamefont{Motruk}},
  \bibinfo{author}{\bibfnamefont{M.~P.} \bibnamefont{Zaletel}},
  \bibinfo{author}{\bibfnamefont{R.~S.~K.} \bibnamefont{Mong}},
  \bibnamefont{and} \bibinfo{author}{\bibfnamefont{F.}~\bibnamefont{Pollmann}},
  \emph{\bibinfo{title}{Density matrix renormalization group on a cylinder in
  mixed real and momentum space}}, \bibinfo{journal}{Phys. Rev. B}
  \textbf{\bibinfo{volume}{93}}, \bibinfo{pages}{155139}
  (\bibinfo{year}{2016}).

\end{thebibliography}


\setcounter{equation}{0}
\renewcommand{\theequation}{S\arabic{equation}}

\setcounter{figure}{0}
\renewcommand{\thefigure}{S\arabic{figure}}

\section{Supplementary material for "Momentum-space entanglement and Loschmidt echo in Luttinger liquids after a quantum quench"}

\section{Non-integrable model}

\begin{figure}[h!]
\centering
\psfrag{x}[t][][1][0]{$J_z/J$}
\psfrag{y}[b][][1][0]{$\color{blue}-\frac{\ln(P_{max,1})}{N}$, \color{red} $-\ln$(largest two overlaps)/$N$}
\includegraphics[width=6.6cm]{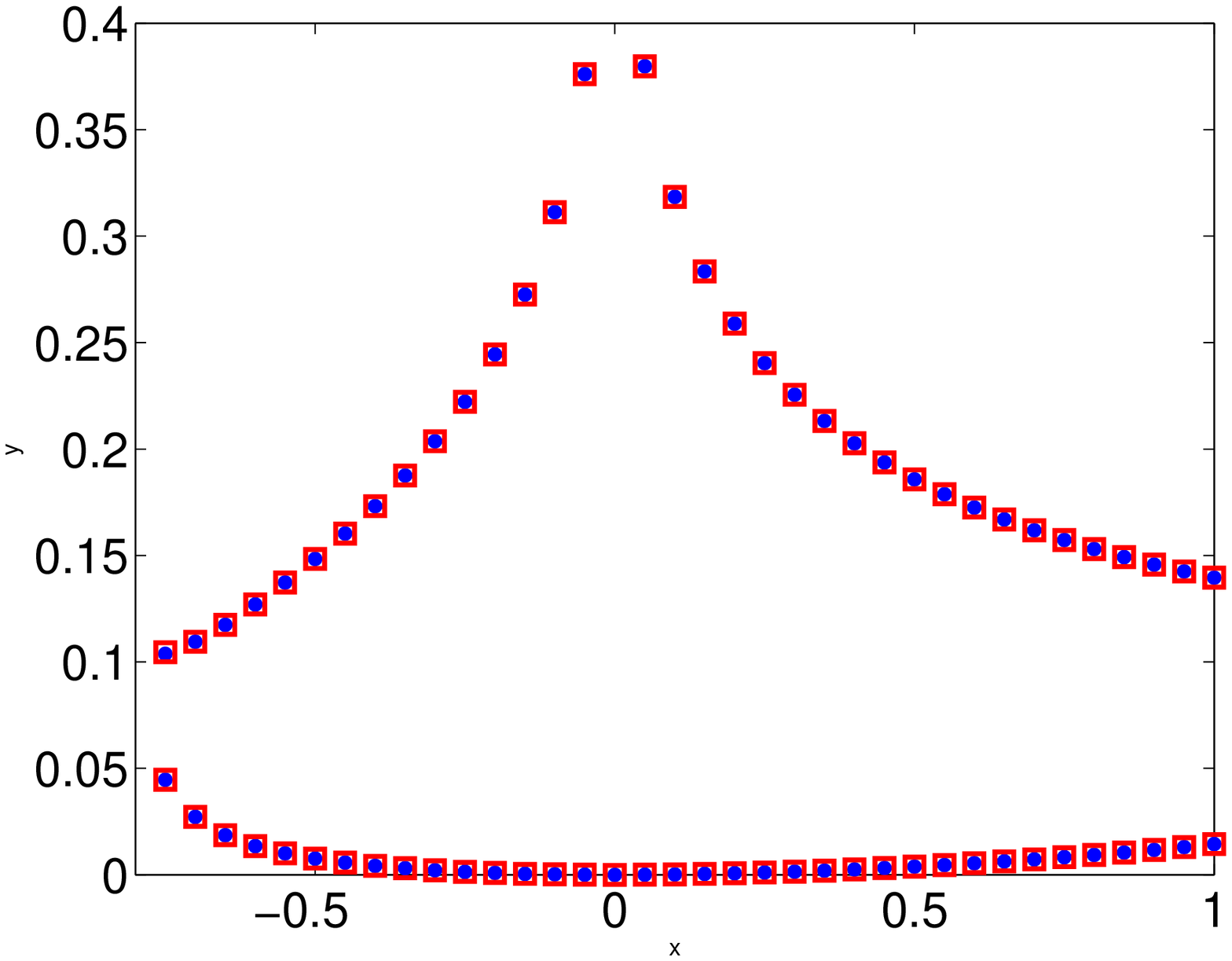}
\caption{The first two largest eigenvalues of the momentum space entanglement spectrum (blue circles) of the non-integrable model in
Eq. \eqref{hnonint} for $N=22$ and $b=1/3$, together with the ground state and excited state overlaps (red squares).}
\label{nonintfig1}
\end{figure}

A non-integrable extension of Eq. (10) in the main text is a 1D spinless fermion model with nearest neighbour interaction and with nearest and \emph{next} nearest neighbour hoppings \cite{mukerjee,bartsch}. 
Its Hamiltonian in momentum space is 
\begin{gather}
H=J\sum_{k}\left(\cos(k)+b\cos(2k)\right)c_k^{\dagger} c_k^{\phantom{\dagger}}+\nonumber\\
+\frac{J_z}{N}\sum_{k,p,q}\cos(q)c^\dagger_{p-q}c^{\phantom{\dagger}}_pc^\dagger_{k+q}c^{\phantom{\dagger}}_k,
\label{hnonint}
\end{gather}
where $c$'s are fermionic annihilation operators in momentum-space, $N$ the number of lattice sites and $k=2\pi m/N$, $n=1\dots N$ for periodic boundary conditions, 
and $Jb$ parameterizes the strength of the second nearest neighbour hopping.
For $b>1/2$ new Fermi points appear at half filling, therefore we 
restrict ourselves to the $0<b<1/2$ regime and choose $b=1/3$ without loss of generality.
Since the model with $b\neq 0$ is non-integrable, no analytic expression is available for the LL parameter, $K$. We numerically diagonalize the Hamiltonian in equilibrium as described in the main text, and the first two largest momentum
space entanglement eigenvalues, together with the first two largest overlaps, are shown in Fig. \ref{nonintfig1}.

\begin{figure}[h!]
\centering
\psfrag{x}[t][][1][0]{$tJ$}
\psfrag{y}[b][][1][0]{$\color{blue}\ln(P_{max})/N$, \color{red} $\ln(\mathcal L(t))/N$}
\psfrag{xx}[t][][0.8][0]{$k/\pi$}
\psfrag{yy}[b][][0.8][0]{$\cos(k)+b\cos(2k)+b$}
\includegraphics[width=6cm]{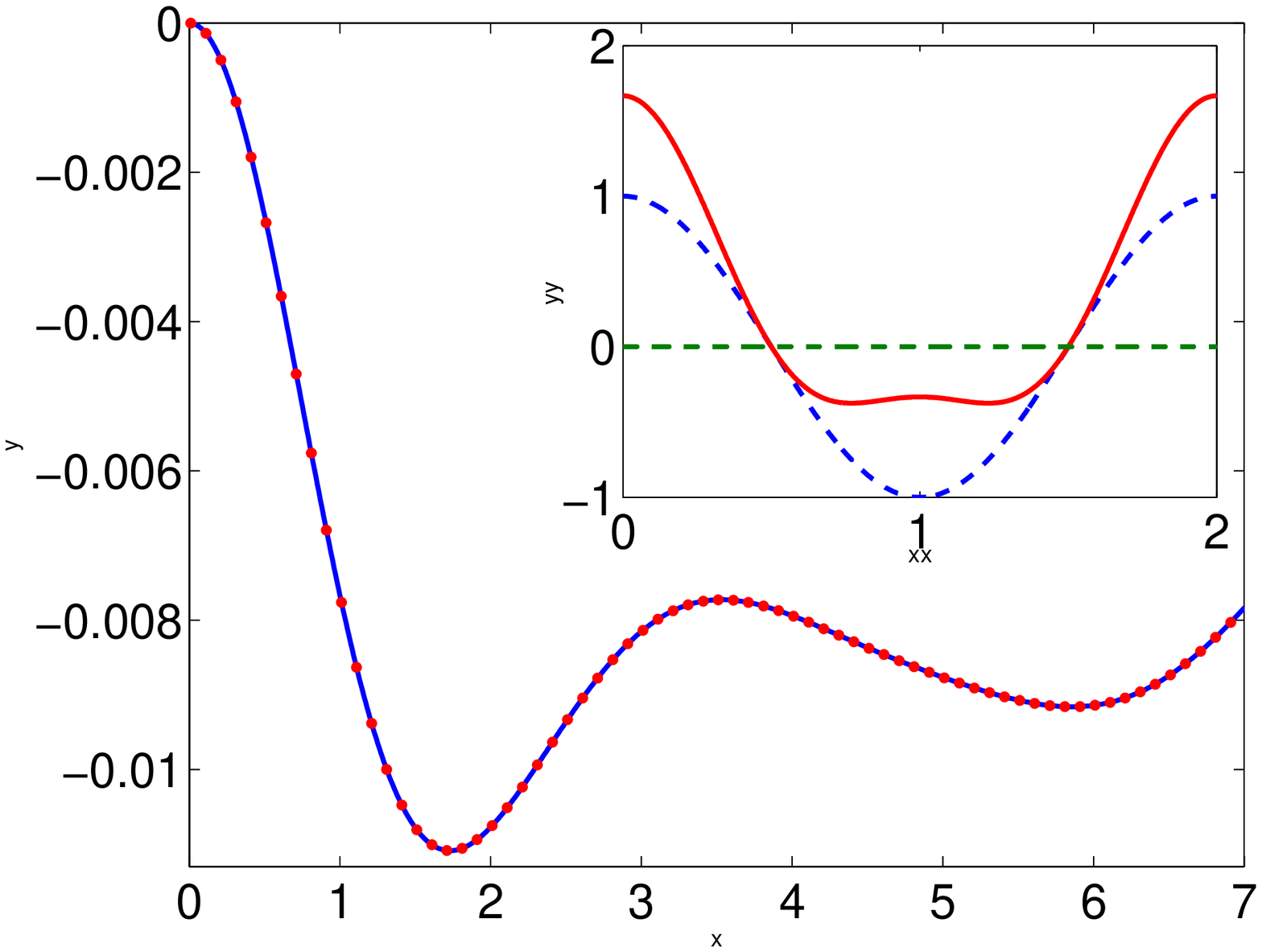}
\caption{The momentum space entanglement ground state energy (blue solid line) and the Loschmidt echo (red dots) are plotted for the non-integrable model 
of Eq. \eqref{hnonint} with $N=22$ and $b=1/3$
after a quench from $J_z=0$ to $J_z=0.5 J$. The inset visualizes the non-interacting  spectrum for $b=0$ (blue dashed line) and $b=1/3$ (red solid line) 
shifted to zero at half filling.}
\label{nonintfig2}
\end{figure}

After an interaction quench from $J_z=0$ to $J_z=0.5J$, the time evolution of the entanglement ground state energy, together with the Loschmidt echo, are shown in Fig. \ref{nonintfig2}. 
As seen in both figures, there is remarkable agreement between the lowest eigenvalues of $\rho_A$ and wavefunction overlaps in the non-integrable case. 
Other values of $b$ and $N$ yield similar agreement between the entanglement eigenvalues and the overlaps.

\section{Antiperiodic boundary conditions}

\begin{figure}[h!]
\centering
\psfrag{x}[t][][1][0]{$J_z/J$}
\psfrag{y}[b][][1][0]{$\color{blue}-\frac{\ln(P_{max,1})}{N}$, \color{red} $-\ln$(largest two overlaps)/$N$}
\includegraphics[width=6.6cm]{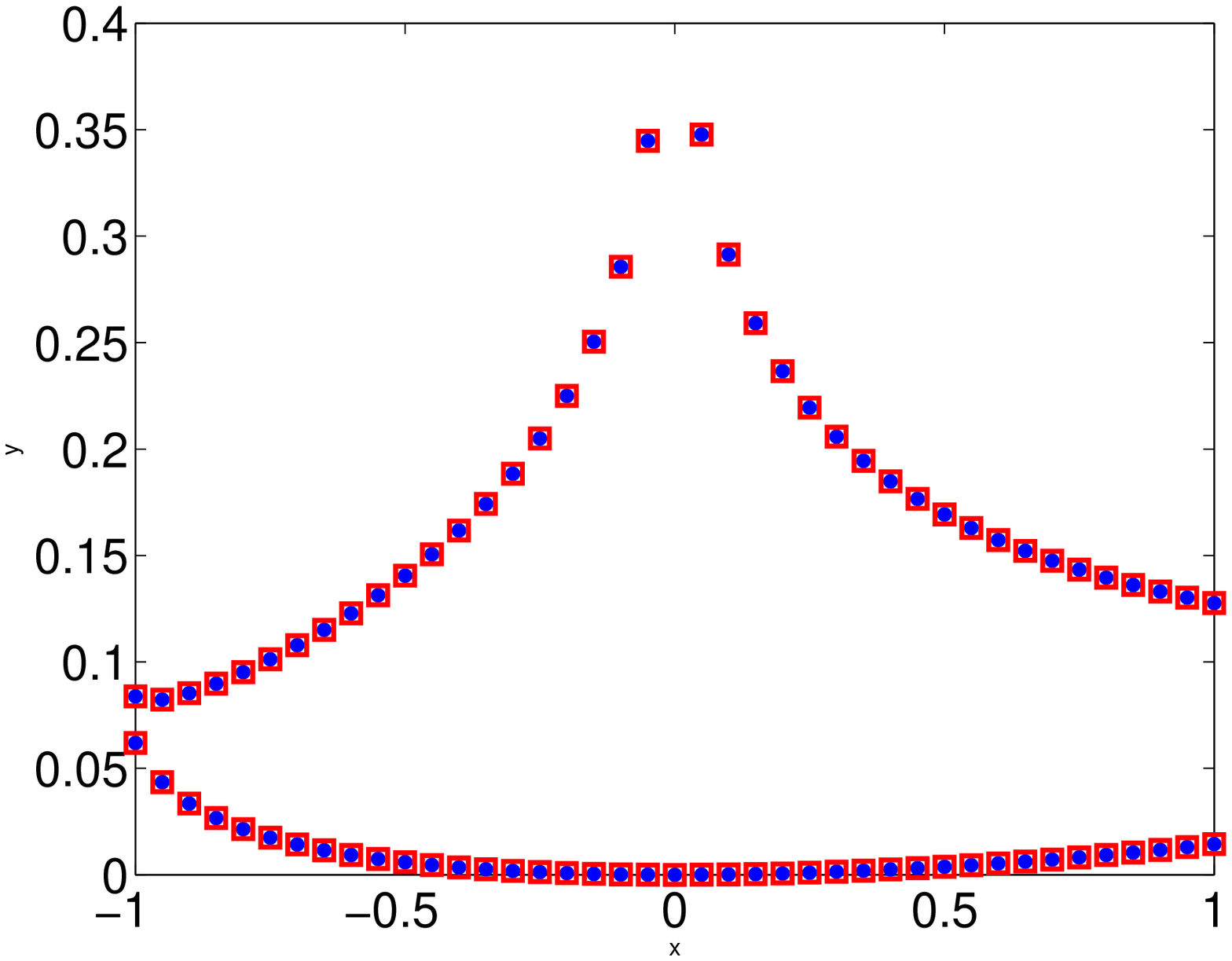}
\caption{The first two largest eigenvalues of the momentum space entanglement spectrum (blue circles) of Eq. \eqref{hnonint} with $b=0$ and $N=24$ with antiperiodic 
boundary conditions, together with the ground state and excited state overlaps (red squares).}
\label{bc}
\end{figure}

\begin{figure}[h!]
\centering
\psfrag{x}[t][][1][0]{$tJ$}
\psfrag{y}[b][][1][0]{$\color{blue}\ln(P_{max})/N$, \color{red} $\ln(\mathcal L(t))/N$}
\includegraphics[width=6.6cm]{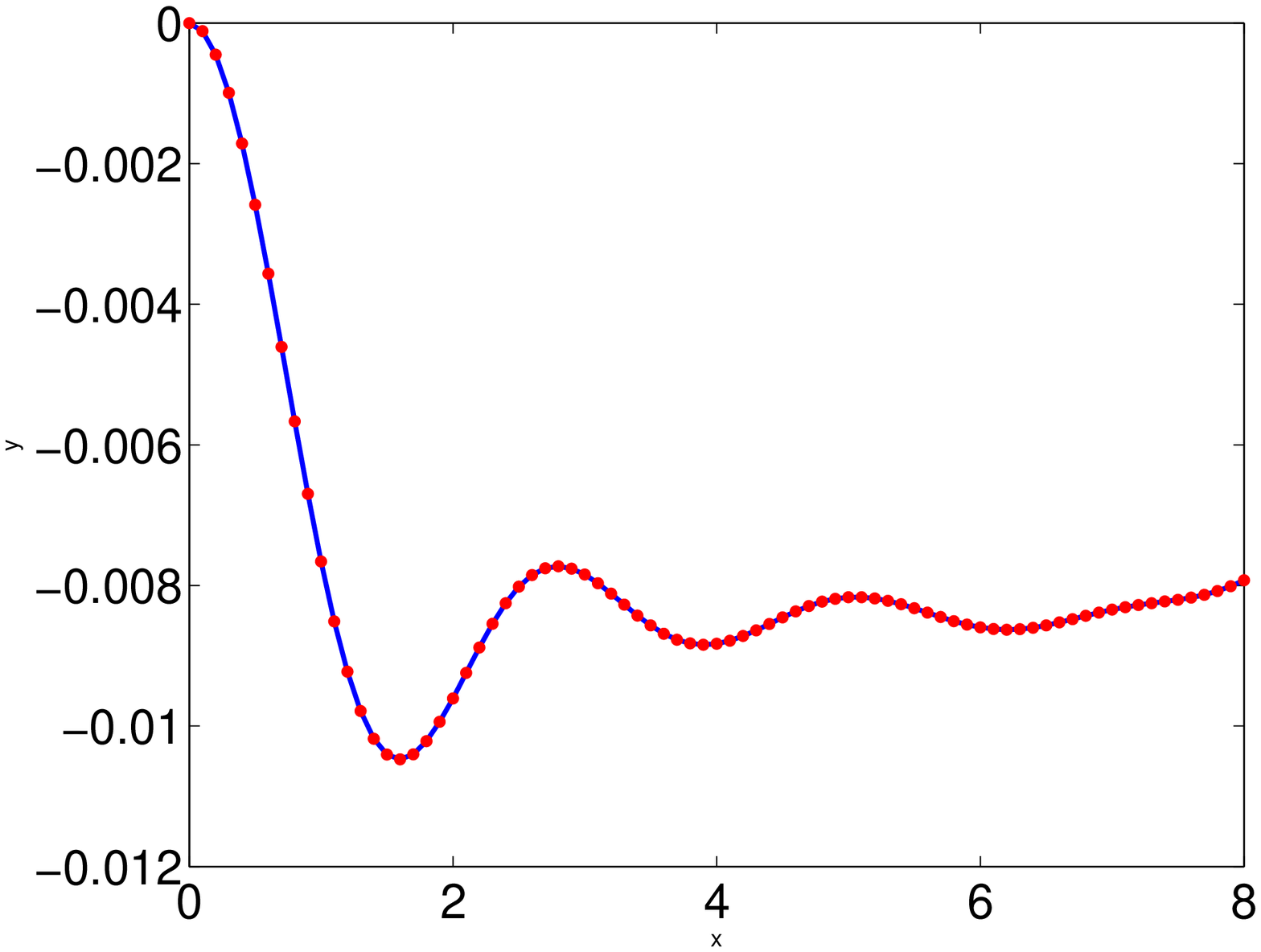}
\caption{The momentum space entanglement ground state energy (blue solid line) and the Loschmidt echo (red dots) are plotted with $b=0$, antiperiodic boundary 
conditions and  $N=24$
after a quench from $J_z=0$ to $J_z=0.5 J$.}
\label{bcquench}
\end{figure}

Finally, we turn to different boundary conditions. Antiperiodic boundary conditions yield $k=(\pi+2\pi m)/N$, $n=1\dots N$. 
To avoid a degenerate Fermi sea in the non-interacting case, we consider the $N=24$ chain and $b=0$ (integrable case). 
The numerical results are plotted in Fig. \ref{bc}. We see excellent agreement between entanglement eigenvalues and overlaps in the presence of antiperiodic boundary conditions. 
We have also tried different twisted boundary conditions with $k=(\theta+2\pi m)/N$ for several values of $0<\theta<1/2$, and system sizes and the agreement was the same. 
After a $J_z$ quench  from $J_z=0$ to $J_z=0.5J$, the time evolution of the entanglement ground state energy and  the Loschmidt echo are visualized  in Fig. \ref{bcquench} 
with remarkable agreement again.

\end{document}